\documentclass[prl,aps,showkeys,showpacs,superscriptaddress,twocolumn,amsmath,amssymb]
{revtex4-1} 
\usepackage{graphics}
\usepackage{bm}
\usepackage{subfigure}
\usepackage{epsfig}
\usepackage[bookmarks,colorlinks,citecolor=blue]{hyperref}
\usepackage{natbib}
\usepackage{siunitx}

\DeclareSIUnit\Molar{\textsc{m}}
\newcommand{\beginsupplement}{%
        \setcounter{table}{0}
        \renewcommand{\thetable}{S\arabic{table}}%
        \setcounter{figure}{0}
        \renewcommand{\thefigure}{S\arabic{figure}}%
     }
\usepackage{xcolor}
\usepackage[normalem]{ulem}

\usepackage{graphicx}	
\begin{document}
\title{Shear thickening and jamming of dense suspensions:
the \emph{roll} of friction
}
\author{Abhinendra Singh}
\email{asingh.iitkgp@gmail.com}
\affiliation{Institute for Molecular Engineering, University of Chicago, Chicago, Illinois 60637, USA}
\affiliation{James Franck Institute, University of Chicago, Chicago, Illinois 60637, USA}
\author{Christopher Ness}
\affiliation{School of Engineering, University of Edinburgh, Edinburgh EH9 3FG, United Kingdom}
\author{Ryohei Seto}
\affiliation{Wenzhou Institute, University of Chinese Academy of Sciences, Wenzhou, Zhejiang 325000, China}
\author{Juan J de Pablo}
\affiliation{Institute for Molecular Engineering, University of Chicago, Chicago, Illinois 60637, USA}
\affiliation{Materials Science Division, Argonne National Laboratory, Lemont, Illinois 60439, USA}
\author{Heinrich M Jaeger}
\affiliation{James Franck Institute, University of Chicago, Chicago, Illinois 60637, USA}
\affiliation{Department of Physics, The University of Chicago, Chicago, Illinois 60637, USA}
\date{\today} 
\begin{abstract}
Particle-based simulations of discontinuous shear thickening (DST) and shear jamming (SJ) suspensions are used to study the role of stress-activated constraints, with an emphasis on resistance to gear-like rolling. Rolling friction decreases the volume fraction required for DST and SJ, in quantitative agreement with real-life suspensions with adhesive surface chemistries and ``rough'' particle shapes. It sets a distinct structure of the frictional force network compared to only sliding friction, and from a dynamical perspective leads to an increase in the velocity correlation length, in part responsible for the increased viscosity. The physics of rolling friction is thus a key element in achieving a comprehensive understanding of strongly shear-thickening materials. 
\end{abstract}
\maketitle

\paragraph{Introduction:}
The flow properties of dense suspensions of non-Brownian particles are critical in numerous natural and industrial processes~\citep{Denn_2018,mewis_colloidal_2011,Brown_2014,Coussot_1997,van2018concrete,blanco2019conching}.
Under shear, such suspensions can display extreme non-Newtonian phenomena~\citep{Brown_2014,Denn_2018,guazzelli2018rheology} that originate in details of interfacial forces~\citep{mewis_colloidal_2011,Singh_2019} as well as \emph{frictional} contact forces~\citep{Comtet_2017,Fernandez_2013}.
In particular, strong shear thickening, a phenomenon of both fundamental interest and practical importance~\citep{Denn_2018,Brown_2014,mewis_colloidal_2011},
represents a crossover from unconstrained to constrained tangential pairwise particle motions as the imposed shear stress $\sigma$ increases 
and a repulsive force threshold ($F_0$, defined below) is exceeded~\citep{Seto_2013a,Mari_2014,Lin_2015,Singh_2018}.
Such \emph{stress-activated} constraints can originate from Coulombic, static friction~\citep{Fernandez_2013, Seto_2013a, Mari_2014, Guy_2015, Lin_2015, Ness_2016,Comtet_2017,Singh_2018}
or from a combined effect of hydrodynamics and asperities~\citep{Jamali_2019}.
Static friction enhances correlated motion and stabilizes load-bearing force networks against buckling, thereby leading to a reduced jamming volume fraction $\phi_J^\mu(\sigma)$~\citep{LiuNagel_AnnRev}
and ultimately an increased viscosity set by $\eta \equiv \sigma/\dot{\gamma} \sim (1-\phi/\phi_J^\mu(\sigma))^{-2}$, with $\dot{\gamma}$ the shear rate.
Indeed, the prevailing theoretical description of shear thickening is 
a two-state model by Wyart and Cates (WC)~\citep{Wyart_2014} that interpolates linearly between
frictionless and frictional $\eta$ divergences as $\sigma$ is increased
using, as a scalar order parameter,
the fraction of contacts that are frictional.
At volume fractions close to $\phi_J^\mu$, $\eta$ can jump by orders of magnitude (discontinuous shear thickening (DST)~\citep{Seto_2013a,Denn_2018,Brown_2014}) upon minuscule changes in $\dot{\gamma}$; at $\phi > \phi_J^\mu$, the suspension can even form a solid-like, shear jammed (SJ) state~\citep{Peters_2016, Singh_2018,Seto_2019,han2019stress}
\footnote{SJ is not expected in the absence of static friction~\citep{Jamali_2019}, however, as fluid-mediated forces vanish upon cessation of flow thus restoring a finite viscosity.}.

An important fundamental question
is how the nature of force transmission changes in the presence of stress-activated particle friction
and, specifically, whether direct contacts constrain both sliding and 
rolling pairwise motion.
The consequences of constraining particle motion by sliding (coefficient $\mu_s$)  and rolling (coefficient $\mu_r$) friction for the rheology and microscopic dynamics during DST and SJ remain largely unexplored,
despite recent works that attest to its importance~\citep{Mari_2019,richards2020role}.

\begin{figure}[tbp]
\centering
{\includegraphics[trim = 0mm 89.4mm 144.7mm 0mm,clip,width=.45\textwidth]{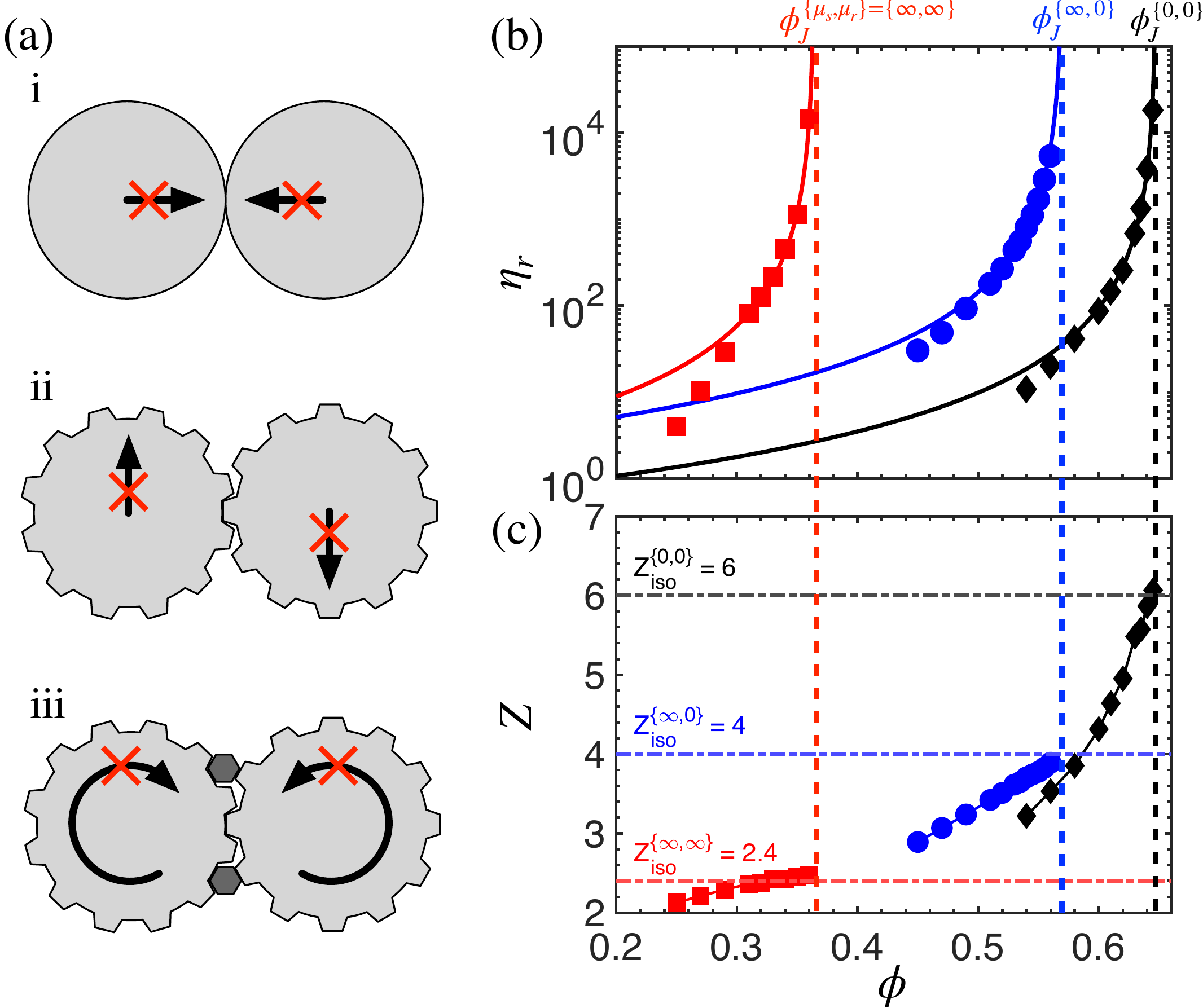}}
\caption{%
Jamming and constraints.
(a) Different types of constraint:
(i) hard sphere, $\{\mu_s, \mu_r\}= \{0, 0\}$;
(ii) infinite sliding friction, $ \{\infty, 0\}$;
(iii) infinite sliding and rolling frictions, $ \{\infty, \infty\}$;
(b,\,c) Simulation data for constraints
(i) (black diamonds);
(ii) (blue circles);
and (iii) (red squares).
(b) Relative viscosity $\eta_r\equiv\eta/\eta_0$\,vs.\,volume fraction $\phi$.
Solid lines are fits to $\eta_r = \bigl(1-\phi/\phi_J^{\{\mu_s,\mu_r\}}\bigr)^{-2}$,
where
$\phi_J^{\{0,0\}} = 0.6477$,
 $\phi_J^{\{\infty,0\}} = 0.5702$,
 and $\phi_J^{\{\infty, \infty\}} = 0.3648$.
(c) Contact number $Z$ (only non-rattler particles)\,vs.\,volume fraction $\phi$.
Horizontal dashed lines indicate the isostatic conditions $Z_{\mathrm{iso}}$; vertical dashed lines are $\phi_J^{\{\mu_s,\mu_r\}}$ used in (b).
 }
\label{figure1}
\end{figure}

In this letter, we address this issue directly and demonstrate the role of constraints
numerically by marrying the physics of both rolling and sliding friction from dry granular materials with a well-established simulation approach
 for shear-thickening suspensions~\citep{Seto_2013a, Mari_2014}.
Sketched in Fig.\,\ref{figure1}\,(a) are schematics of pairwise contacts
illustrating
hard-sphere (i),
sliding (ii),
and rolling (iii) constraints.
When particles experience a hard-sphere constraint only, but no friction,
 $\eta$ diverges when $Z$, the number of non-rattler contacts per particle~\citep{Seto_2019}, equals its so-called isostatic value $Z_{\mathrm{iso}}^{\{\mu_s = 0, \mu_r =0\}}=6$. This occurs at a specific $\phi_J^{\{\mu_s=0,\mu_r=0\}}$~\citep{OHern_2003,LiuNagel_AnnRev}, which in our 3D simulation for a bidisperse suspension is $\approx0.65$.
The constraints offered by friction at contact confer enhanced mechanical stability
so that $\eta$ can diverge for $Z<6$ and $\phi_J<0.65$.
For instance, large $\mu_s$ leads to $Z_{\mathrm{iso}}^{\{\infty,0\}}=4$ at $\phi_J^{\{\infty,0\}}\approx0.57$~\citep{Hecke_2009}.
Incorporating both rolling and sliding friction further lowers the limiting number of contacts to $Z_{\mathrm{iso}}^{\{\infty,\infty \}} = D(D+1)/(2D-1) = 2.4$ (in 3D)~\citep{Mari_2019}~\footnote{glassy systems with covalent bonds also report a limit of 2.4 when bending is constrained~\citep{he1985}},
so that $\phi_J^{\{\infty,\infty\}}\approx 0.36$ (see Figs.\,\ref{figure1}\,(b) and (c)).
This simple argument already demonstrates that the viscosity is highly sensitive to the nature of tangential constraints.

Most natural and industrially relevant suspensions, including cornstarch--water mixtures, an archetypical shear-thickening suspension,
comprise faceted particles with asperities (and, in some cases, adhesive interactions originating from surface chemistry)~\citep{Lootens_2005,Hsiao_2017,Hsu_2018,james2018interparticle}.
Such features lead to interlocking between particle surfaces introducing new physics not describable by sliding friction alone, suggesting that resistance to rolling is important.
Moreover, in the dry granular literature it has been shown that a direct consequence of angular particle shape is hindered particle rotation, and that the rheology can be reproduced by incorporating rolling friction along with sliding friction~\citep{Estrada_2011,ai2011assessment}. 
Meanwhile, in dry tribology adhesive forces between particles are known to resist rolling due to flattening of the contact point~\citep{Dominik_1995,Marshall_2014}.
Recent suspension studies have demonstrated that short-ranged particle-particle interactions such as hydrogen bonding may similarly not only increase sliding friction but also introduce a small amount of weak, reversible adhesion~\citep{james2018interparticle,James_2019}. The latter can lead to stress-activated rolling friction.
Crucially, such suspensions exhibit DST at $\phi\lessapprox0.45$~\citep{Lootens_2005,Neuville_2012,Hsiao_2017,Hsu_2018,hsiao2019experimental},
whereas simulations that include only sliding friction consistently report the lower bound for DST as $\phi\approx0.56$~\citep{Ness_2016,Singh_2018}.
This dramatic discrepancy impedes quantitative prediction of experimental behavior despite recent advances in the field~\citep{Denn_2018}.

The physics of \emph{stress-activated} rolling friction is thus
an attractive candidate to account for the longstanding disparity 
between experiments and simulations:
it is micromechanically well-motivated as it captures the effect of facets, asperities and surface chemistry;
it can, on the grounds of constraint counting, account for the low-$\phi$ SJ observed experimentally; and
it is consistent with the WC model~\cite{Wyart_2014,Singh_2018,Guy_2018}\footnote{Indeed the theory with rolling friction was recently discussed in \citep{Mari_2019})}.

\paragraph{Method:}
We simulate a bidisperse suspension, an equal volume fraction mixture of $2000$ inertialess spheres of radii $a$ and $1.4a$,
suspended in a density-matched Newtonian fluid of viscosity $\eta_0$.
Under imposed shear stress $\sigma_{xy}$ (referred to as $\sigma$ below and described in \citep{Seto_2019})
the suspension flows with time-dependent shear rate $\dot\gamma(t)$ in a 3D Lees--Edwards periodic domain.
After omitting the start-up flow transient (which typically lasts  $\mathcal{O}(1)$ strain units)
we report $\eta_r \equiv {\sigma}/\eta_0\langle \dot{\gamma}\rangle$, where angle brackets imply time average over the steady state.
The particles are subject to Stokes drag and interact through short-range 
pairwise hydrodynamic lubrication interactions $\boldsymbol{F}_H$ (see~\citep{Mari_2014}),
repulsive forces $\boldsymbol{F}_R$, and contact interactions $\boldsymbol{F}_C$. 
The repulsive force acts normally and decays with interparticle surface separation $h$ 
over a Debye length $\lambda$ as $|\boldsymbol{F}_R|  =  F_0 \exp(-h/\lambda)$ (we use $\lambda=0.01a$).
This gives rise to a stress scale $\sigma_0 \equiv F_0/6\pi a^2$,
related by an $\mathcal{O}(1)$ prefactor (which may very weakly depend on $\lambda$) to the crossover from lubricated, frictionless contacts between particles to direct, frictional ones.
The contact interaction is modeled using linear springs~\citep{Mari_2014},
incorporating both sliding and rolling friction 
using the algorithm described by~\citet{Luding_2008}.
Contacts obey Coulomb's friction law for both sliding and rolling modes: $|F_{C,t}| \le \mu_s |F_{C,n}|$ and $|F_{C,r}| \le \mu_r |F_{C,n}|$.
Rolling friction introduces a resistance to motion that is not a force but a torque. Thus, the rolling friction force $|F_{C,r}|$,
which is proportional to the relative rolling displacement,
is a quasi-force that does not contribute to the force balance 
and is calculated only to compute the rolling torque.
Hindered rolling motion leads to contacting particles that, 
under compression, must rotate as a solid body as though glued to each other.
Under tension, meanwhile, contacts simply break.
Further details are available in the Supplemental Materials~\citep{NoteX}.

\paragraph{Overview of bulk rheology results:}

\begin{figure}
\centering
\includegraphics[trim = 0mm 161.5mm 94.3mm 0mm, clip,width=.48\textwidth,page=1]{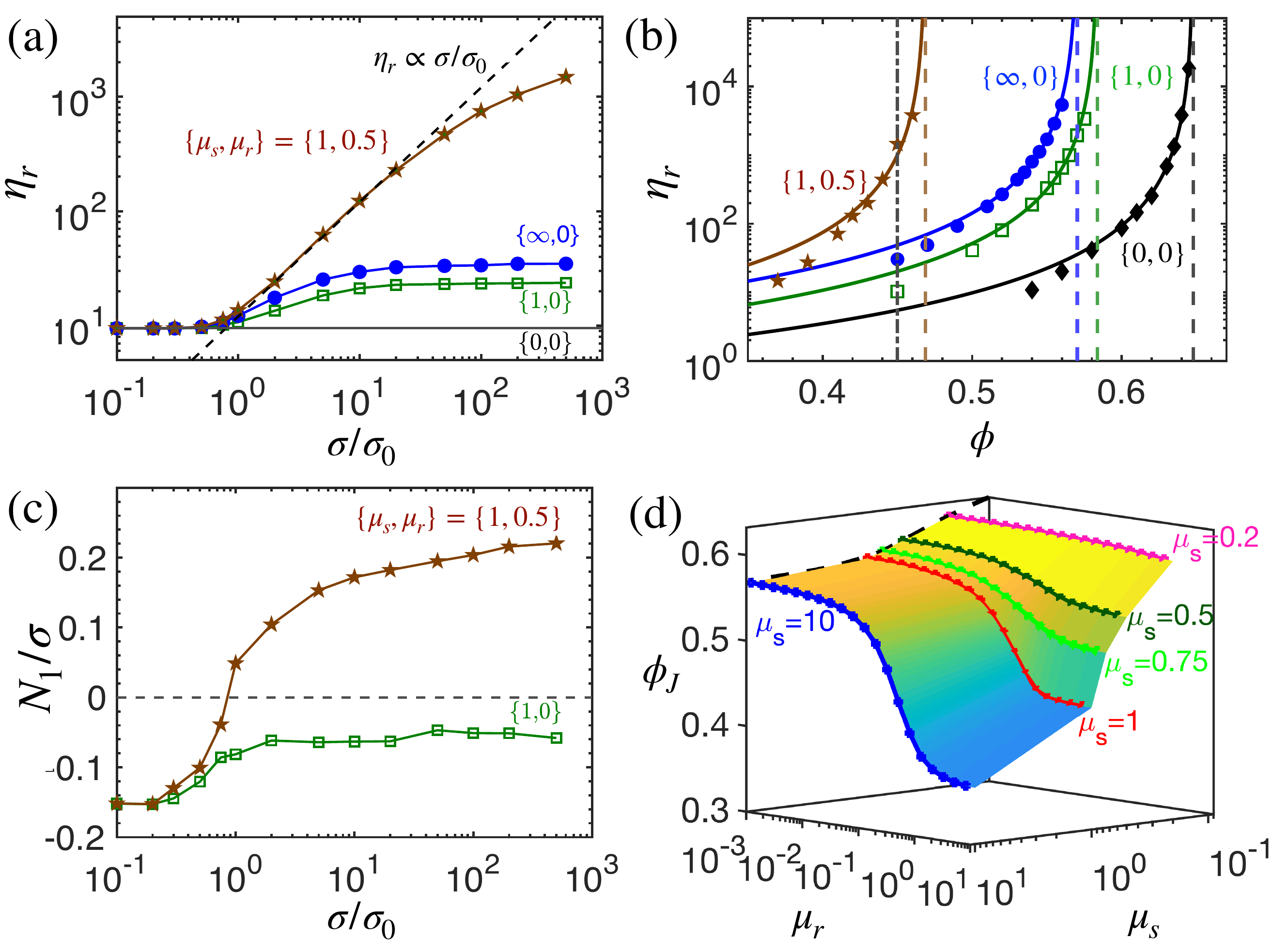}
\caption{%
Rheology with rolling friction.
(a) Viscosity $\eta_{\mathrm{r}}$ as a function of shear stress ${\sigma}/{\sigma}_0$ at $\phi=0.45$.
Weak shear thickening is observed in simulations with sliding friction only 
($\{\mu_s, \mu_r\}=\{1,0\}$ and $\{\infty,0\}$),
whereas those with rolling friction 
($\{\mu_s, \mu_r\}=\{1,0.5\}$) display DST ($\eta_r \propto \sigma/\sigma_0$, dashed line) at the same $\phi$.
(b) $\eta_r$ divergence with $\phi$ for \{$\mu_s$, $\mu_r$\} pairs in (a).
(c) The ratio $N_1/\sigma$ of the first normal stress difference to the shear stress
as a function of $\sigma/\sigma_0$ at $\phi=0.45$.
$N_1/\sigma$ with siding friction only ($\{\mu_s, \mu_r\}=\{1,0\}$) remains small and negative,
while with rolling friction ($\{\mu_s, \mu_r\}=\{1,0.5\}$) it turns positive.
(d) 
$\phi_J$ dependence on sliding $\mu_s$ and rolling $\mu_r$ friction coefficients.
The dashed line is the dependence of $\phi_{J}$ on $\mu_s$ with $\mu_r=0$.
Solid lines indicate the dependence of $\phi_J$ on $\mu_r$ for several values of $\mu_s$.
 }
\label{figure2}
\end{figure}

Shown in Fig.\,\ref{figure2}\,(a) is the relative viscosity $\eta_r$ as a function of scaled shear stress $\sigma/\sigma_0$
for three combinations of friction coefficients 
\{$\mu_s$, $\mu_r$\} 
at $\phi=0.45$.
Setting $\mu_r=0$ at this $\phi$ leads to continuous shear thickening (CST) regardless of the value of $\mu_s$, 
whereas $\mu_r>0$ leads to DST as evidenced by $\eta_r \propto \sigma/\sigma_0$ (dashed line).
Because frictional contacts are stress-activated (as also assumed by the WC model), at
$\sigma/\sigma_0 \ll1$, $\eta_r$ resides on 
the \{0, 0\} branch of Fig.\,\ref{figure2}\,(b)
(squares and line).
Increasing $\sigma/\sigma_0$ at fixed $\phi$, $\eta_r$ transitions  
to a frictional branch
as direct contacts appear.
The extent of shear thickening is set simply by the position of $\phi_J^{\{0,0\}}$
relative to $\phi_J^{\{\mu_s,\mu_r\}}$:
the more constraints are added, the lower $\phi_J$ becomes
and the more severe shear thickening is.
Thus incorporating rolling friction recovers the surprisingly low SJ volume fraction $\phi=0.45$ (for these parameters) observed experimentally in the case of suspensions with rough particles~\citep{Lootens_2005, Hsu_2018, Hsiao_2017}.
Recent theory~\citep{Mari_2019} suggests a generalization of the WC model
to reflect more selective force transmission due to rolling friction,
causing a wider range of stress over which thickening occurs.
This is consistent with our findings and also the experimental observations of \citet{Hsu_2018}.
Figure\,\ref{figure2}\,(c) shows 
the ratio $N_1/\sigma$ (with $N_1 \equiv \sigma_{xx}-\sigma_{yy}$) of the first normal stress difference,
indicating the reorientation angle of the eigenvectors of the stress, 
for $\{\mu_s, \mu_r\} = \{1, 0\}$ (squares) and
$\{1, 0.5\}$ (stars) at $\phi=0.45$. 
Simulations without rolling friction exhibit a small, negative $N_1/\sigma$ 
for the entire range of $\sigma$~\citep{Mari_2014},
while simulations with rolling friction exhibit a sign change to positive values;
they are even larger than the reported result 
near jamming without rolling friction~\citep{Seto_2018}.
Our results indicate that the contact network can behave more elastic-like 
due to more stable contacts with rolling friction.
This is consistent with recent experiments on tunable rough particles~\citep{Hsu_2018, Hsiao_2017} that showed a similar transition in $N_1$ upon increasing particle roughness.

In Fig.\,\ref{figure2}\,(d), we present a comprehensive map of $\phi_J^{\{\mu_s,\mu_r\}}$,
generated by simulating the limit $\sigma/\sigma_0 \to \infty$ (by setting $F_0=0$) for a broad range of $\mu_s$, $\mu_r$ and $\phi$,
and extracting $\phi_J$ by fitting the viscosity to 
$\eta_r = \bigl(1-\phi/\phi_J^{\{\mu_s,\mu_r\}} \bigr)^{-2}$.
For every value of $\mu_s$,
we observe that $\phi_J$ decreases with increasing $\mu_r$.
For the lowest $\mu_s=0.2$ simulated here 
the effect of rolling friction on $\phi_J$ is rather modest.
With increasing $\mu_s$ the dependence of $\phi_J$ on $\mu_r$ becomes stronger and we observe saturation at $\mu_r \ge 1$.
Especially interesting is the $\mu_s$ range between 0.5 and 1, where $\phi_J$ decreases rapidly as small amounts of rolling friction come into play.
%
For the case of $\mu_s \to \infty$, the change in $\mu_r$ from $10^{-3}$ to 10 decreases $\phi_J$ from 0.57 to 0.36. 
Our results suggest that for suspensions with small sliding friction coefficient ($\mu_s \le 0.2$)
$\phi_J$ is independent of $\mu_r$.
Meanwhile for particles with higher sliding friction, $\mu_s \ge 0.35$,
rolling constraints can drastically affect the rheological behavior.

\paragraph{Comparison with experiments:}
\begin{figure}
\centering
%
{\includegraphics[trim = 0mm 162.5mm 104.3mm 0mm, clip,width=.48\textwidth,page=1]{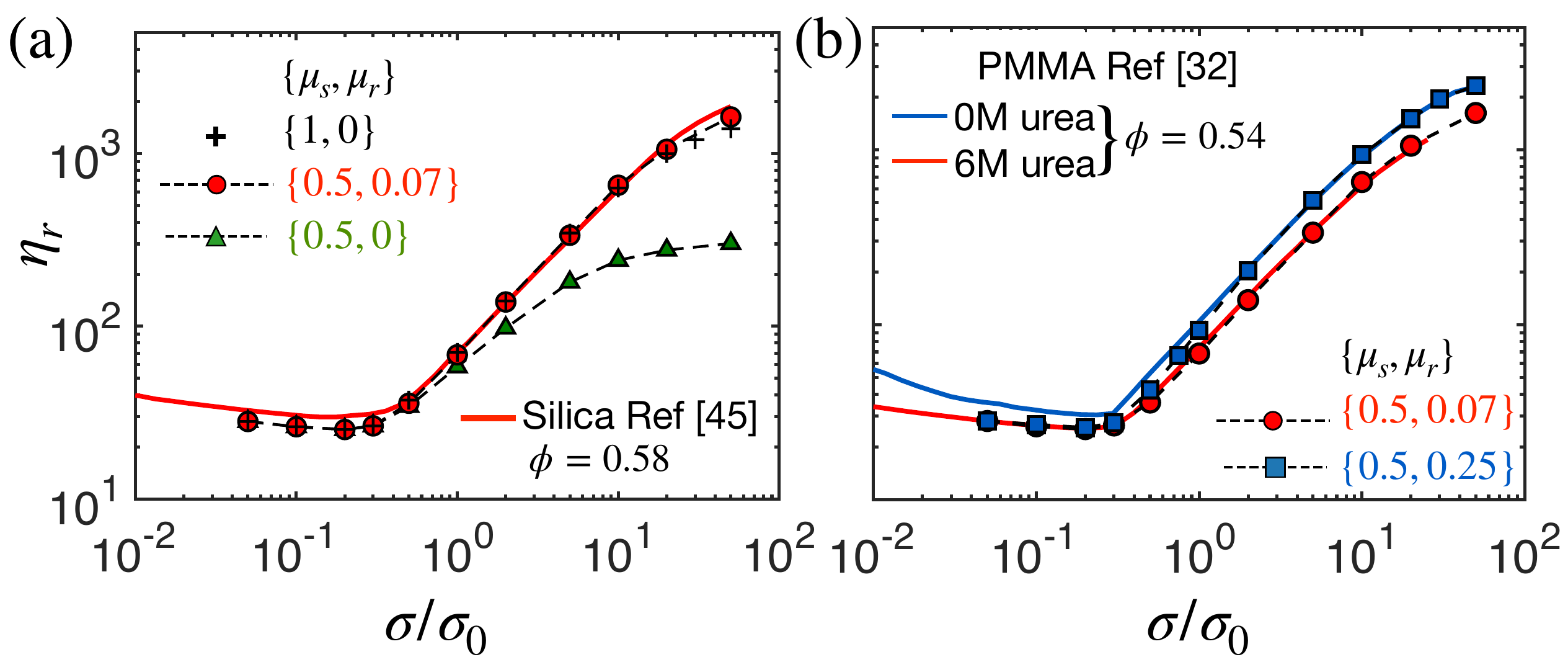}}
\caption{%
Comparison with experiments.
(a) Experimental data from \citep{Royer_2016} (line) for $\phi=0.58$ and simulation data for $\phi=0.56$ (symbols) for 
various combinations of $\mu_s$ and $\mu_r$. To scale the experimental results we use $F_0= \SI{1.5}{\nano\newton}$.
(b) Experimental data from \citep{james2018interparticle} (lines) for $\phi=0.54$ and simulation data for $\phi=0.56$ (symbols). 
To scale the experimental results we use $F_0= \SI{1}{\nano\newton}$ and $F_0= \SI{0.3}{\nano\newton}$ for \SI{6}{\Molar} and \SI{0}{\Molar}, respectively.
 }
\label{figure3}
\end{figure}
Given that the shape of normalized rheological flow curves as in Fig.\,\ref{figure2}\,(a) is controlled by $\phi_J$ as the only free parameter,
any \{$\mu_s$, $\mu_r$\} pair residing 
on a constant-$\phi_J$ contour of appropriate magnitude could fit the experimental data equally well.
Still, there are considerations regarding the magnitude of $\mu_s$.
To reproduce DST seen in experiments with nominally smooth spheres, previous simulations~\citep{Seto_2013a, Mari_2014, Ness_2016, Singh_2018} that only constrained sliding required
$\mu_s  \approx \! 1$.
This is a concern~\citep{Tanner_2016,Denn_2018}, since direct measurements 
typically report 
$\mu_s \lessapprox0.5$~\citep{Tanner_2016,Comtet_2017,james2018interparticle}.
However, from Fig.\,\ref{figure2}\,(d) we find that an equally good fit should be obtainable by reducing $\mu_s$ to 0.5 and adding some rolling friction, around 1/10 of $\mu_s$.
We demonstrate this  in Fig.\,\ref{figure3}\,(a) for silica spheres with data from Royer \emph{et al.}~\cite{Royer_2016}, which are reproduced very well  using the pair \{$\mu_s$, $\mu_r$\} = \{$0.5$, 0.07\}, at large stresses possibly even better than by $\{1, 0\}$ \footnote{If $N_1$ data are also available, we may be able to determine $\mu_s$ and $\mu_r$ uniquely.}.
While small, this rolling resistance is important to capture the physics of frictional particle-particle interactions: \{$0.5$, 0\}  underpredicts the viscosity significantly.
 
We next consider experiments by \citet{james2018interparticle}, in which hydrogen bonding between surface-functionalized PMMA/ITA spheres in an aqueous solvent was shown to increase the \emph{effective} interparticle friction (Fig.\,\ref{figure3}\,(b); we scaled the two curves such that the onset stress for shear thickening is the same and coincides with the simulation data, i.e., $\sigma/\sigma_0=0.3$).
When hydrogen bonding is suppressed by adding \SI{6}{\Molar} urea, the PMMA particles behave similar to other smooth spheres at comparable $\phi$.
Consequently, the same  $\{0.5,\,0.07\}$ pair as in Fig.\,\ref{figure3}\,(a) reproduces the data very well (as would have $ \{1,\,0\}$).
Without urea, hydrogen bonding is operative and introduces a measurable `stickiness’~\citep{James_2019}  to the contact force.
Figure\,\ref{figure3}\,(b) shows that this additional adhesion can be modelled well by increasing the rolling resistance from $\mu_r=0.07$ to $\mu_r=0.25$.
 
For particles with very rough surfaces that can geometrically interlock a large $\mu_s  \approx \! 1$ may be appropriate.  
In experiments with such particles, Hsu \emph{et al.} found that $\phi_J$ dropped as low as 0.44~\citep{Hsu_2018}. Figure\,\ref{figure2}\,(d) indicates that sliding friction by itself cannot produce such small $\phi_J$, implying additional rolling constraints. 
Indeed, by dialing up both $\mu_s$ and $\mu_r$ to values near 1 we can closely mimic the reduction in $\phi_J$ seen by Hsu \emph{et al.}~\citep{Hsu_2018}
\paragraph{Microstructural behavior:}
%
\begin{figure}[tbp]
\centering
{\includegraphics[trim = 0mm 45mm 105mm 0mm, clip,width=.48\textwidth,page=1]{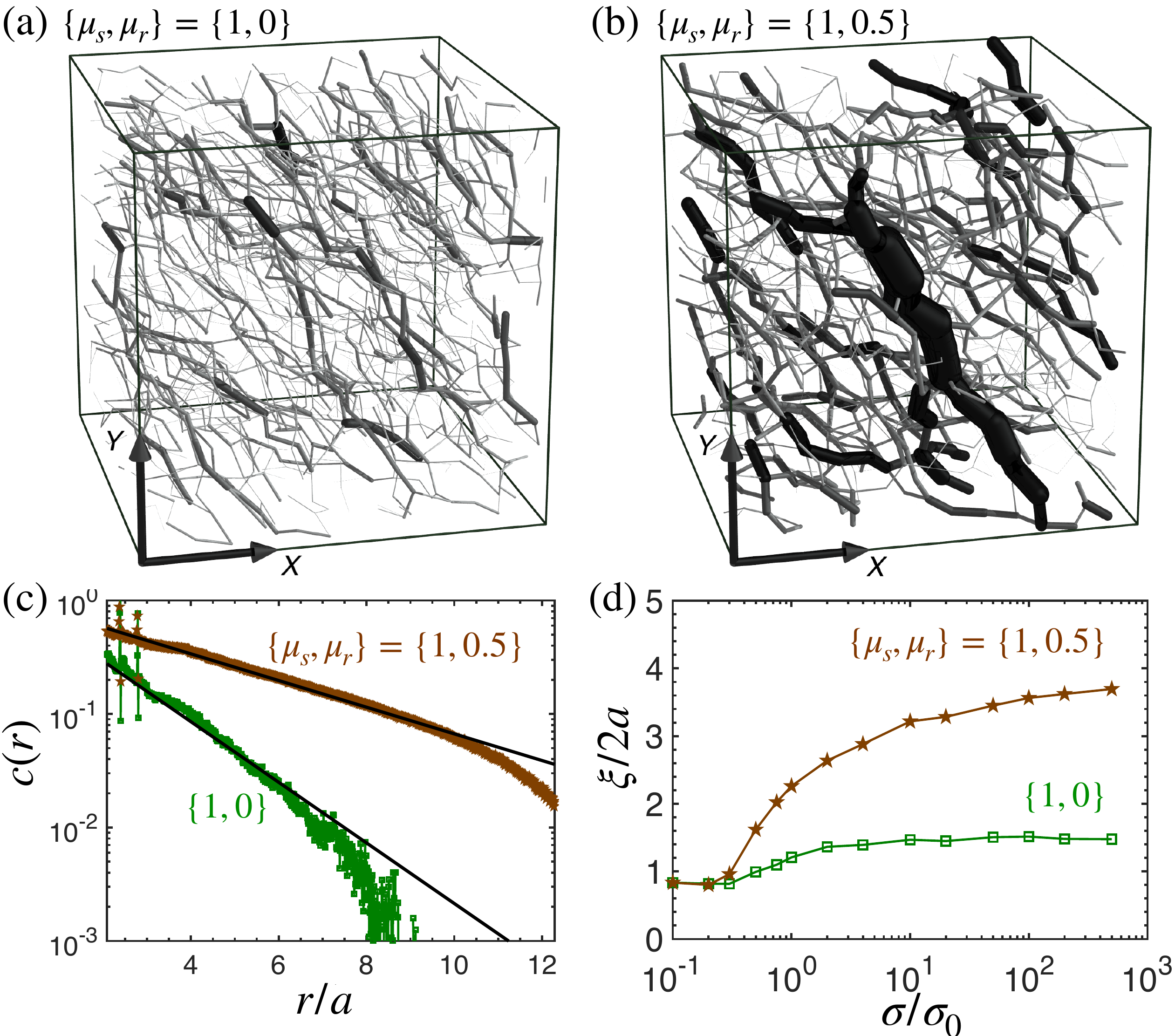}}
\caption{%
Microstructural consequences of increased rolling friction.
(a,\,b) 
Force network snapshots in the steady state for $\phi = 0.45$ 
 at ${\sigma/\sigma_0}=500$ for $\{\mu_s, \mu_r\} = \{1, 0\}$ (a) and $\{1, 0.5\}$ (b). 
 The width and darkness of the line segments represent the contact force magnitude.
(c) Non-affine velocity correlation function $c(r)$ plotted against $r/L_z$ for 
$\{\mu_s, \mu_r \}$ = \{1, 0.5\} and \{1, 0\} 
at $\sigma/\sigma_0=500$.
Lines correspond to  $c(r) = \alpha \exp(-r/\xi)$. 
(d) Velocity correlation length $\xi/2a$ with and without rolling friction,
 as a function of $\sigma/\sigma_0$ for $\phi=0.45$.
 }
\label{figure4}
\end{figure}
We finally address the microscopic underpinnings for the differences in the measured viscosity
with and without rolling friction,
focussing on the force network formed by frictional contacts
and the correlation of the fluctuating non-affine velocity.

Figures\,\ref{figure4}\,(a) and (b)
compare the stress transmission patterns
with $\mu_r=0$ and $\mu_r = 0.5$ at $\phi = 0.45$.
The line segments indicate frictional contacts.
Note that all contact force network structures are transient,
continuously flowing, breaking and re-forming under the bulk shearing motion.
The force networks shown in Fig.\,\ref{figure4} are for $\sigma/\sigma_0=500$, for which $\eta_r$ differs by almost 2 orders of magnitude, see Fig.\,\ref{figure2}\,(a).
The frictional forces 
appear as roughly linear structures (force chains) along the compression axis, i.e., 
along $y = -x$~\citep{Mari_2014, gameiro2019interaction}.
Comparing Figs.~\ref{figure4}(a) and (b), 
force transmission in the presence of rolling friction is much more spatially localized and directed than with only sliding friction.
Indeed in the former case the force chains are thicker and darker, carrying larger force compared to the latter.
In the case without rolling friction, the force chains easily buckle and rearrange under shear.
 However, by constraining the rolling mode buckling is suppressed, so chains can more robustly prevail under applied stress.
Hence, the particles exhibit less relative movement with respect to their neighbors and show enhanced correlation.

The velocity correlation quantifies this collective motion, as used previously for dry granular particles~\citep{lois2007spatial}. Here we define it similar to~\citep{Ness_2016}:
\begin{equation}
c(r) \equiv \frac{\sum \limits_i\sum \limits_{j>i} \bar{\boldsymbol{v}}_i \cdot \bar{\boldsymbol{v}}_j \delta(|r_{ij}| -r) }{\sum \limits_i\sum \limits_{j>i} \delta(|r_{ij}| -r) }~,
\end{equation}
where $ \bar{\boldsymbol{v}}_i$ and $ \bar{\boldsymbol{v}}_j$ are the fluctuating velocity vectors that are averaged over a time interval corresponding to approximately a single particle displacement due to mean flow.
Figure~\ref{figure4}\,(c) displays $c(r)$ for $\sigma/\sigma_0=500$, demonstrating the enhancement of the velocity correlations in the case with rolling friction compared to that without rolling friction.
We find that $c(r)$ decays approximately exponentially with the distance between particle centers $r$.
The correlation length $\xi$ that can be extracted from fits of data as in Fig.~\ref{figure4}\,(c) 
 to $c(r) = \alpha \exp(-r/\xi)$
as a function of stress $\sigma$ 
is displayed in Fig.\,\ref{figure4}\,(d).
We find that the correlation length increases with stress, implying the correlated motion increases with $\sigma$, but that sliding friction alone shows only a very mild increase.
On the other hand, simulations with additional rolling friction show a significant  increase in the correlation length.
The implied difference observed in the rheology due to the enhanced collective motion
of particles  can also be observed directly in videos based on the simulations (see~\citep{NoteX}).

\paragraph{Conclusions:}
We have studied the rheology of dense suspensions interacting through 
short-range lubrication and contact interactions with stress-activated sliding and rolling friction.
The latter generates a constraint  on relative particle movement
 that allows us to reproduce experimental features 
  including $\phi_J<0.5$.
Inhibited rolling means that particles must move or gyrate together as a temporal (but not permanent) cluster, confirmed by the enhanced velocity correlation, which is in part
responsible for the increased viscosity.
When only sliding motion is constrained, the load-bearing force chains need orthogonal support to avoid buckling~\citep{Radjai_1998}. By contrast, constraining both rolling and sliding motions leads to a more anisotropic force chain structure that can sustain external loads unaided, leading to a lower jamming point.
The rolling friction in this work is intended to capture any particle-scale effects that hinder rolling, whether they originate from physical surface properties such as shape and roughness~\citep{Hsiao_2017,Hsu_2018} or surface chemistry~\citep{james2018interparticle}; more sophisticated models will be required to make quantitative predictions for more complex particle shapes~\cite{cwalina2016rheology,cwalina2017rheology}.

\begin{acknowledgments}
We appreciate stimulating discussions with Grayson Jackson, Nicole James and Mike van der Naald. 
We acknowledge Romain Mari for co-developing simulation codes 
to implement rolling friction.
AS, JJdP and HJ acknowledge support from the Center for Hierarchical Materials Design (CHiMaD) under award number 70NANB19H005 (US Dept. Commerce) and from the Chicago MRSEC, which is supported by NSF DMR-1420709.
HJ acknowledges additional support from the Army Research Office under grants W911NF-16-1-0078 and W911NF-19-1-0245. 
CN acknowledges support from the Royal Academy of Engineering under the Research Fellowship scheme. 
\end{acknowledgments}

\bibliography{dst} 

\begin{thebibliography}{55}%
\makeatletter
\providecommand \@ifxundefined [1]{%
 \@ifx{#1\undefined}
}%
\providecommand \@ifnum [1]{%
 \ifnum #1\expandafter \@firstoftwo
 \else \expandafter \@secondoftwo
 \fi
}%
\providecommand \@ifx [1]{%
 \ifx #1\expandafter \@firstoftwo
 \else \expandafter \@secondoftwo
 \fi
}%
\providecommand \natexlab [1]{#1}%
\providecommand \enquote  [1]{``#1''}%
\providecommand \bibnamefont  [1]{#1}%
\providecommand \bibfnamefont [1]{#1}%
\providecommand \citenamefont [1]{#1}%
\providecommand \href@noop [0]{\@secondoftwo}%
\providecommand \href [0]{\begingroup \@sanitize@url \@href}%
\providecommand \@href[1]{\@@startlink{#1}\@@href}%
\providecommand \@@href[1]{\endgroup#1\@@endlink}%
\providecommand \@sanitize@url [0]{\catcode `\\12\catcode `\$12\catcode
  `\&12\catcode `\#12\catcode `\^12\catcode `\_12\catcode `\%12\relax}%
\providecommand \@@startlink[1]{}%
\providecommand \@@endlink[0]{}%
\providecommand \url  [0]{\begingroup\@sanitize@url \@url }%
\providecommand \@url [1]{\endgroup\@href {#1}{\urlprefix }}%
\providecommand \urlprefix  [0]{URL }%
\providecommand \Eprint [0]{\href }%
\providecommand \doibase [0]{http://dx.doi.org/}%
\providecommand \selectlanguage [0]{\@gobble}%
\providecommand \bibinfo  [0]{\@secondoftwo}%
\providecommand \bibfield  [0]{\@secondoftwo}%
\providecommand \translation [1]{[#1]}%
\providecommand \BibitemOpen [0]{}%
\providecommand \bibitemStop [0]{}%
\providecommand \bibitemNoStop [0]{.\EOS\space}%
\providecommand \EOS [0]{\spacefactor3000\relax}%
\providecommand \BibitemShut  [1]{\csname bibitem#1\endcsname}%
\let\auto@bib@innerbib\@empty
\bibitem [{\citenamefont {Denn}\ \emph {et~al.}(2018)\citenamefont {Denn},
  \citenamefont {Morris},\ and\ \citenamefont {Bonn}}]{Denn_2018}%
  \BibitemOpen
  \bibfield  {author} {\bibinfo {author} {\bibfnamefont {M.~M.}\ \bibnamefont
  {Denn}}, \bibinfo {author} {\bibfnamefont {J.~F.}\ \bibnamefont {Morris}}, \
  and\ \bibinfo {author} {\bibfnamefont {D.}~\bibnamefont {Bonn}},\ }\href@noop
  {} {\bibfield  {journal} {\bibinfo  {journal} {Soft Matter}\ }\textbf
  {\bibinfo {volume} {14}},\ \bibinfo {pages} {170} (\bibinfo {year}
  {2018})}\BibitemShut {NoStop}%
\bibitem [{\citenamefont {Mewis}\ and\ \citenamefont
  {Wagner}(2011)}]{mewis_colloidal_2011}%
  \BibitemOpen
  \bibfield  {author} {\bibinfo {author} {\bibfnamefont {J.}~\bibnamefont
  {Mewis}}\ and\ \bibinfo {author} {\bibfnamefont {N.~J.}\ \bibnamefont
  {Wagner}},\ }\href@noop {} {\emph {\bibinfo {title} {Colloidal {Suspension}
  {Rheology}}}}\ (\bibinfo  {publisher} {Cambridge University Press},\ \bibinfo
  {year} {2011})\BibitemShut {NoStop}%
\bibitem [{\citenamefont {Brown}\ and\ \citenamefont
  {Jaeger}(2014)}]{Brown_2014}%
  \BibitemOpen
  \bibfield  {author} {\bibinfo {author} {\bibfnamefont {E.}~\bibnamefont
  {Brown}}\ and\ \bibinfo {author} {\bibfnamefont {H.~M.}\ \bibnamefont
  {Jaeger}},\ }\href@noop {} {\bibfield  {journal} {\bibinfo  {journal} {Rep.
  Prog. Phys.}\ }\textbf {\bibinfo {volume} {77}},\ \bibinfo {pages} {046602}
  (\bibinfo {year} {2014})}\BibitemShut {NoStop}%
\bibitem [{\citenamefont {Coussot}(1997)}]{Coussot_1997}%
  \BibitemOpen
  \bibfield  {author} {\bibinfo {author} {\bibfnamefont {P.}~\bibnamefont
  {Coussot}},\ }\href@noop {} {\emph {\bibinfo {title} {Mudflow Rheology and
  Dynamics}}}\ (\bibinfo  {publisher} {CRC Press},\ \bibinfo {year}
  {1997})\BibitemShut {NoStop}%
\bibitem [{\citenamefont {Van~Damme}(2018)}]{van2018concrete}%
  \BibitemOpen
  \bibfield  {author} {\bibinfo {author} {\bibfnamefont {H.}~\bibnamefont
  {Van~Damme}},\ }\href@noop {} {\bibfield  {journal} {\bibinfo  {journal}
  {Cem. Concr. Res.}\ }\textbf {\bibinfo {volume} {112}},\ \bibinfo {pages} {5}
  (\bibinfo {year} {2018})}\BibitemShut {NoStop}%
\bibitem [{\citenamefont {Blanco}\ \emph {et~al.}(2019)\citenamefont {Blanco},
  \citenamefont {Hodgson}, \citenamefont {Hermes}, \citenamefont {Besseling},
  \citenamefont {Hunter}, \citenamefont {Chaikin}, \citenamefont {Cates},
  \citenamefont {Van~Damme},\ and\ \citenamefont {Poon}}]{blanco2019conching}%
  \BibitemOpen
  \bibfield  {author} {\bibinfo {author} {\bibfnamefont {E.}~\bibnamefont
  {Blanco}}, \bibinfo {author} {\bibfnamefont {D.~J.}\ \bibnamefont {Hodgson}},
  \bibinfo {author} {\bibfnamefont {M.}~\bibnamefont {Hermes}}, \bibinfo
  {author} {\bibfnamefont {R.}~\bibnamefont {Besseling}}, \bibinfo {author}
  {\bibfnamefont {G.~L.}\ \bibnamefont {Hunter}}, \bibinfo {author}
  {\bibfnamefont {P.~M.}\ \bibnamefont {Chaikin}}, \bibinfo {author}
  {\bibfnamefont {M.~E.}\ \bibnamefont {Cates}}, \bibinfo {author}
  {\bibfnamefont {I.}~\bibnamefont {Van~Damme}}, \ and\ \bibinfo {author}
  {\bibfnamefont {W.~C.~K.}\ \bibnamefont {Poon}},\ }\href@noop {} {\bibfield
  {journal} {\bibinfo  {journal} {Proc. Natl. Acad. Sci. USA}\ }\textbf
  {\bibinfo {volume} {116}},\ \bibinfo {pages} {10303} (\bibinfo {year}
  {2019})}\BibitemShut {NoStop}%
\bibitem [{\citenamefont {Guazzelli}\ and\ \citenamefont
  {Pouliquen}(2018)}]{guazzelli2018rheology}%
  \BibitemOpen
  \bibfield  {author} {\bibinfo {author} {\bibfnamefont {{\'E}.}~\bibnamefont
  {Guazzelli}}\ and\ \bibinfo {author} {\bibfnamefont {O.}~\bibnamefont
  {Pouliquen}},\ }\href@noop {} {\bibfield  {journal} {\bibinfo  {journal} {J.
  Fluid Mech.}\ }\textbf {\bibinfo {volume} {852}} (\bibinfo {year}
  {2018})}\BibitemShut {NoStop}%
\bibitem [{\citenamefont {Singh}\ \emph {et~al.}(2019)\citenamefont {Singh},
  \citenamefont {Pednekar}, \citenamefont {Chun}, \citenamefont {Denn},\ and\
  \citenamefont {Morris}}]{Singh_2019}%
  \BibitemOpen
  \bibfield  {author} {\bibinfo {author} {\bibfnamefont {A.}~\bibnamefont
  {Singh}}, \bibinfo {author} {\bibfnamefont {S.}~\bibnamefont {Pednekar}},
  \bibinfo {author} {\bibfnamefont {J.}~\bibnamefont {Chun}}, \bibinfo {author}
  {\bibfnamefont {M.~M.}\ \bibnamefont {Denn}}, \ and\ \bibinfo {author}
  {\bibfnamefont {J.~F.}\ \bibnamefont {Morris}},\ }\href@noop {} {\bibfield
  {journal} {\bibinfo  {journal} {Phys. Rev. Lett.}\ }\textbf {\bibinfo
  {volume} {122}},\ \bibinfo {pages} {098004} (\bibinfo {year}
  {2019})}\BibitemShut {NoStop}%
\bibitem [{\citenamefont {Comtet}\ \emph {et~al.}(2017)\citenamefont {Comtet},
  \citenamefont {Chatt{\'e}}, \citenamefont {Nigu{\`e}s}, \citenamefont
  {Bocquet}, \citenamefont {Siria},\ and\ \citenamefont {Colin}}]{Comtet_2017}%
  \BibitemOpen
  \bibfield  {author} {\bibinfo {author} {\bibfnamefont {J.}~\bibnamefont
  {Comtet}}, \bibinfo {author} {\bibfnamefont {G.}~\bibnamefont {Chatt{\'e}}},
  \bibinfo {author} {\bibfnamefont {A.}~\bibnamefont {Nigu{\`e}s}}, \bibinfo
  {author} {\bibfnamefont {L.}~\bibnamefont {Bocquet}}, \bibinfo {author}
  {\bibfnamefont {A.}~\bibnamefont {Siria}}, \ and\ \bibinfo {author}
  {\bibfnamefont {A.}~\bibnamefont {Colin}},\ }\href@noop {} {\bibfield
  {journal} {\bibinfo  {journal} {Nat. Comm.}\ }\textbf {\bibinfo {volume}
  {8}},\ \bibinfo {pages} {15633} (\bibinfo {year} {2017})}\BibitemShut
  {NoStop}%
\bibitem [{\citenamefont {Fernandez}\ \emph {et~al.}(2013)\citenamefont
  {Fernandez}, \citenamefont {Mani}, \citenamefont {Rinaldi}, \citenamefont
  {Kadau}, \citenamefont {Mosquet}, \citenamefont {Lombois-Burger},
  \citenamefont {Cayer-Barrioz}, \citenamefont {Herrmann}, \citenamefont
  {Spencer},\ and\ \citenamefont {Isa}}]{Fernandez_2013}%
  \BibitemOpen
  \bibfield  {author} {\bibinfo {author} {\bibfnamefont {N.}~\bibnamefont
  {Fernandez}}, \bibinfo {author} {\bibfnamefont {R.}~\bibnamefont {Mani}},
  \bibinfo {author} {\bibfnamefont {D.}~\bibnamefont {Rinaldi}}, \bibinfo
  {author} {\bibfnamefont {D.}~\bibnamefont {Kadau}}, \bibinfo {author}
  {\bibfnamefont {M.}~\bibnamefont {Mosquet}}, \bibinfo {author} {\bibfnamefont
  {H.}~\bibnamefont {Lombois-Burger}}, \bibinfo {author} {\bibfnamefont
  {J.}~\bibnamefont {Cayer-Barrioz}}, \bibinfo {author} {\bibfnamefont {H.~J.}\
  \bibnamefont {Herrmann}}, \bibinfo {author} {\bibfnamefont {N.~D.}\
  \bibnamefont {Spencer}}, \ and\ \bibinfo {author} {\bibfnamefont
  {L.}~\bibnamefont {Isa}},\ }\href@noop {} {\bibfield  {journal} {\bibinfo
  {journal} {Phys. Rev. Lett.}\ }\textbf {\bibinfo {volume} {111}},\ \bibinfo
  {pages} {108301} (\bibinfo {year} {2013})}\BibitemShut {NoStop}%
\bibitem [{\citenamefont {Seto}\ \emph {et~al.}(2013)\citenamefont {Seto},
  \citenamefont {Mari}, \citenamefont {Morris},\ and\ \citenamefont
  {Denn}}]{Seto_2013a}%
  \BibitemOpen
  \bibfield  {author} {\bibinfo {author} {\bibfnamefont {R.}~\bibnamefont
  {Seto}}, \bibinfo {author} {\bibfnamefont {R.}~\bibnamefont {Mari}}, \bibinfo
  {author} {\bibfnamefont {J.~F.}\ \bibnamefont {Morris}}, \ and\ \bibinfo
  {author} {\bibfnamefont {M.~M.}\ \bibnamefont {Denn}},\ }\href@noop {}
  {\bibfield  {journal} {\bibinfo  {journal} {Phys. Rev. Lett.}\ }\textbf
  {\bibinfo {volume} {111}},\ \bibinfo {pages} {218301} (\bibinfo {year}
  {2013})}\BibitemShut {NoStop}%
\bibitem [{\citenamefont {Mari}\ \emph {et~al.}(2014)\citenamefont {Mari},
  \citenamefont {Seto}, \citenamefont {Morris},\ and\ \citenamefont
  {Denn}}]{Mari_2014}%
  \BibitemOpen
  \bibfield  {author} {\bibinfo {author} {\bibfnamefont {R.}~\bibnamefont
  {Mari}}, \bibinfo {author} {\bibfnamefont {R.}~\bibnamefont {Seto}}, \bibinfo
  {author} {\bibfnamefont {J.~F.}\ \bibnamefont {Morris}}, \ and\ \bibinfo
  {author} {\bibfnamefont {M.~M.}\ \bibnamefont {Denn}},\ }\href@noop {}
  {\bibfield  {journal} {\bibinfo  {journal} {J. Rheol.}\ }\textbf {\bibinfo
  {volume} {58}},\ \bibinfo {pages} {1693} (\bibinfo {year}
  {2014})}\BibitemShut {NoStop}%
\bibitem [{\citenamefont {Lin}\ \emph {et~al.}(2015)\citenamefont {Lin},
  \citenamefont {Guy}, \citenamefont {Hermes}, \citenamefont {Ness},
  \citenamefont {Sun}, \citenamefont {Poon},\ and\ \citenamefont
  {Cohen}}]{Lin_2015}%
  \BibitemOpen
  \bibfield  {author} {\bibinfo {author} {\bibfnamefont {N.~Y.~C.}\
  \bibnamefont {Lin}}, \bibinfo {author} {\bibfnamefont {B.~M.}\ \bibnamefont
  {Guy}}, \bibinfo {author} {\bibfnamefont {M.}~\bibnamefont {Hermes}},
  \bibinfo {author} {\bibfnamefont {C.}~\bibnamefont {Ness}}, \bibinfo {author}
  {\bibfnamefont {J.}~\bibnamefont {Sun}}, \bibinfo {author} {\bibfnamefont
  {W.~C.~K.}\ \bibnamefont {Poon}}, \ and\ \bibinfo {author} {\bibfnamefont
  {I.}~\bibnamefont {Cohen}},\ }\href@noop {} {\bibfield  {journal} {\bibinfo
  {journal} {Phys. Rev. Lett.}\ }\textbf {\bibinfo {volume} {115}},\ \bibinfo
  {pages} {228304} (\bibinfo {year} {2015})}\BibitemShut {NoStop}%
\bibitem [{\citenamefont {Singh}\ \emph {et~al.}(2018)\citenamefont {Singh},
  \citenamefont {Mari}, \citenamefont {Denn},\ and\ \citenamefont
  {Morris}}]{Singh_2018}%
  \BibitemOpen
  \bibfield  {author} {\bibinfo {author} {\bibfnamefont {A.}~\bibnamefont
  {Singh}}, \bibinfo {author} {\bibfnamefont {R.}~\bibnamefont {Mari}},
  \bibinfo {author} {\bibfnamefont {M.~M.}\ \bibnamefont {Denn}}, \ and\
  \bibinfo {author} {\bibfnamefont {J.~F.}\ \bibnamefont {Morris}},\
  }\href@noop {} {\bibfield  {journal} {\bibinfo  {journal} {J. Rheol.}\
  }\textbf {\bibinfo {volume} {62}},\ \bibinfo {pages} {457} (\bibinfo {year}
  {2018})}\BibitemShut {NoStop}%
\bibitem [{\citenamefont {Guy}\ \emph {et~al.}(2015)\citenamefont {Guy},
  \citenamefont {Hermes},\ and\ \citenamefont {Poon}}]{Guy_2015}%
  \BibitemOpen
  \bibfield  {author} {\bibinfo {author} {\bibfnamefont {B.~M.}\ \bibnamefont
  {Guy}}, \bibinfo {author} {\bibfnamefont {M.}~\bibnamefont {Hermes}}, \ and\
  \bibinfo {author} {\bibfnamefont {W.~C.~K.}\ \bibnamefont {Poon}},\
  }\href@noop {} {\bibfield  {journal} {\bibinfo  {journal} {Phys. Rev. Lett.}\
  }\textbf {\bibinfo {volume} {115}},\ \bibinfo {pages} {088304} (\bibinfo
  {year} {2015})}\BibitemShut {NoStop}%
\bibitem [{\citenamefont {Ness}\ and\ \citenamefont {Sun}(2016)}]{Ness_2016}%
  \BibitemOpen
  \bibfield  {author} {\bibinfo {author} {\bibfnamefont {C.}~\bibnamefont
  {Ness}}\ and\ \bibinfo {author} {\bibfnamefont {J.}~\bibnamefont {Sun}},\
  }\href@noop {} {\bibfield  {journal} {\bibinfo  {journal} {Soft Matter}\
  }\textbf {\bibinfo {volume} {12}},\ \bibinfo {pages} {914} (\bibinfo {year}
  {2016})}\BibitemShut {NoStop}%
\bibitem [{\citenamefont {Jamali}\ and\ \citenamefont
  {Brady}(2019)}]{Jamali_2019}%
  \BibitemOpen
  \bibfield  {author} {\bibinfo {author} {\bibfnamefont {S.}~\bibnamefont
  {Jamali}}\ and\ \bibinfo {author} {\bibfnamefont {J.~F.}\ \bibnamefont
  {Brady}},\ }\href@noop {} {\bibfield  {journal} {\bibinfo  {journal} {Phys.
  Rev. Lett.}\ }\textbf {\bibinfo {volume} {123}},\ \bibinfo {pages} {138002}
  (\bibinfo {year} {2019})}\BibitemShut {NoStop}%
\bibitem [{\citenamefont {Liu}\ and\ \citenamefont
  {Nagel}(2010)}]{LiuNagel_AnnRev}%
  \BibitemOpen
  \bibfield  {author} {\bibinfo {author} {\bibfnamefont {A.~J.}\ \bibnamefont
  {Liu}}\ and\ \bibinfo {author} {\bibfnamefont {S.~R.}\ \bibnamefont
  {Nagel}},\ }\href@noop {} {\bibfield  {journal} {\bibinfo  {journal} {Annu.
  Rev. Condens. Matter Phys.}\ }\textbf {\bibinfo {volume} {1}},\ \bibinfo
  {pages} {347} (\bibinfo {year} {2010})}\BibitemShut {NoStop}%
\bibitem [{\citenamefont {Wyart}\ and\ \citenamefont
  {Cates}(2014)}]{Wyart_2014}%
  \BibitemOpen
  \bibfield  {author} {\bibinfo {author} {\bibfnamefont {M.}~\bibnamefont
  {Wyart}}\ and\ \bibinfo {author} {\bibfnamefont {M.~E.}\ \bibnamefont
  {Cates}},\ }\href@noop {} {\bibfield  {journal} {\bibinfo  {journal} {Phys.
  Rev. Lett.}\ }\textbf {\bibinfo {volume} {112}},\ \bibinfo {pages} {098302}
  (\bibinfo {year} {2014})}\BibitemShut {NoStop}%
\bibitem [{\citenamefont {Peters}\ \emph {et~al.}(2016)\citenamefont {Peters},
  \citenamefont {Majumdar},\ and\ \citenamefont {Jaeger}}]{Peters_2016}%
  \BibitemOpen
  \bibfield  {author} {\bibinfo {author} {\bibfnamefont {I.~R.}\ \bibnamefont
  {Peters}}, \bibinfo {author} {\bibfnamefont {S.}~\bibnamefont {Majumdar}}, \
  and\ \bibinfo {author} {\bibfnamefont {H.~M.}\ \bibnamefont {Jaeger}},\
  }\href@noop {} {\bibfield  {journal} {\bibinfo  {journal} {Nature}\ }\textbf
  {\bibinfo {volume} {532}},\ \bibinfo {pages} {214} (\bibinfo {year}
  {2016})}\BibitemShut {NoStop}%
\bibitem [{\citenamefont {Seto}\ \emph {et~al.}(2019)\citenamefont {Seto},
  \citenamefont {Singh}, \citenamefont {Chakraborty}, \citenamefont {Denn},\
  and\ \citenamefont {Morris}}]{Seto_2019}%
  \BibitemOpen
  \bibfield  {author} {\bibinfo {author} {\bibfnamefont {R.}~\bibnamefont
  {Seto}}, \bibinfo {author} {\bibfnamefont {A.}~\bibnamefont {Singh}},
  \bibinfo {author} {\bibfnamefont {B.}~\bibnamefont {Chakraborty}}, \bibinfo
  {author} {\bibfnamefont {M.~M.}\ \bibnamefont {Denn}}, \ and\ \bibinfo
  {author} {\bibfnamefont {J.~F.}\ \bibnamefont {Morris}},\ }\href@noop {}
  {\bibfield  {journal} {\bibinfo  {journal} {Gran. Matt.}\ }\textbf {\bibinfo
  {volume} {21}},\ \bibinfo {pages} {82} (\bibinfo {year} {2019})}\BibitemShut
  {NoStop}%
\bibitem [{\citenamefont {Han}\ \emph {et~al.}(2019)\citenamefont {Han},
  \citenamefont {James},\ and\ \citenamefont {Jaeger}}]{han2019stress}%
  \BibitemOpen
  \bibfield  {author} {\bibinfo {author} {\bibfnamefont {E.}~\bibnamefont
  {Han}}, \bibinfo {author} {\bibfnamefont {N.~M.}\ \bibnamefont {James}}, \
  and\ \bibinfo {author} {\bibfnamefont {H.~M.}\ \bibnamefont {Jaeger}},\
  }\href@noop {} {\bibfield  {journal} {\bibinfo  {journal} {Phys. Rev. Lett.}\
  }\textbf {\bibinfo {volume} {123}},\ \bibinfo {pages} {248002} (\bibinfo
  {year} {2019})}\BibitemShut {NoStop}%
\bibitem [{Note1()}]{Note1}%
  \BibitemOpen
  \bibinfo {note} {SJ is not expected in the absence of static
  friction~\protect \citep {Jamali_2019}, however, as fluid-mediated forces
  vanish upon cessation of flow thus restoring a finite viscosity.}\BibitemShut
  {Stop}%
\bibitem [{\citenamefont {Mari}\ and\ \citenamefont {Seto}(2019)}]{Mari_2019}%
  \BibitemOpen
  \bibfield  {author} {\bibinfo {author} {\bibfnamefont {R.}~\bibnamefont
  {Mari}}\ and\ \bibinfo {author} {\bibfnamefont {R.}~\bibnamefont {Seto}},\
  }\href@noop {} {\bibfield  {journal} {\bibinfo  {journal} {Soft Matter}\
  }\textbf {\bibinfo {volume} {15}},\ \bibinfo {pages} {6650} (\bibinfo {year}
  {2019})}\BibitemShut {NoStop}%
\bibitem [{\citenamefont {Richards}\ \emph {et~al.}(2020)\citenamefont
  {Richards}, \citenamefont {Guy}, \citenamefont {Blanco}, \citenamefont
  {Hermes}, \citenamefont {Poy},\ and\ \citenamefont
  {Poon}}]{richards2020role}%
  \BibitemOpen
  \bibfield  {author} {\bibinfo {author} {\bibfnamefont {J.~A.}\ \bibnamefont
  {Richards}}, \bibinfo {author} {\bibfnamefont {B.~M.}\ \bibnamefont {Guy}},
  \bibinfo {author} {\bibfnamefont {E.}~\bibnamefont {Blanco}}, \bibinfo
  {author} {\bibfnamefont {M.}~\bibnamefont {Hermes}}, \bibinfo {author}
  {\bibfnamefont {G.}~\bibnamefont {Poy}}, \ and\ \bibinfo {author}
  {\bibfnamefont {W.~C.}\ \bibnamefont {Poon}},\ }\href@noop {} {\bibfield
  {journal} {\bibinfo  {journal} {Journal of Rheology}\ }\textbf {\bibinfo
  {volume} {64}},\ \bibinfo {pages} {405} (\bibinfo {year} {2020})}\BibitemShut
  {NoStop}%
\bibitem [{\citenamefont {O'Hern}\ \emph {et~al.}(2003)\citenamefont {O'Hern},
  \citenamefont {Silbert}, \citenamefont {Liu},\ and\ \citenamefont
  {Nagel}}]{OHern_2003}%
  \BibitemOpen
  \bibfield  {author} {\bibinfo {author} {\bibfnamefont {C.~S.}\ \bibnamefont
  {O'Hern}}, \bibinfo {author} {\bibfnamefont {L.~E.}\ \bibnamefont {Silbert}},
  \bibinfo {author} {\bibfnamefont {A.~J.}\ \bibnamefont {Liu}}, \ and\
  \bibinfo {author} {\bibfnamefont {S.~R.}\ \bibnamefont {Nagel}},\ }\href@noop
  {} {\bibfield  {journal} {\bibinfo  {journal} {Phys. Rev. E}\ }\textbf
  {\bibinfo {volume} {68}},\ \bibinfo {pages} {011306} (\bibinfo {year}
  {2003})}\BibitemShut {NoStop}%
\bibitem [{\citenamefont {van Hecke}(2009)}]{Hecke_2009}%
  \BibitemOpen
  \bibfield  {author} {\bibinfo {author} {\bibfnamefont {M.}~\bibnamefont {van
  Hecke}},\ }\href@noop {} {\bibfield  {journal} {\bibinfo  {journal} {J. Phys.
  Condens. Matter}\ }\textbf {\bibinfo {volume} {22}},\ \bibinfo {pages}
  {033101} (\bibinfo {year} {2009})}\BibitemShut {NoStop}%
\bibitem [{Note2()}]{Note2}%
  \BibitemOpen
  \bibinfo {note} {Glassy systems with covalent bonds also report a limit of
  2.4 when bending is constrained~\protect \citep {he1985}}\BibitemShut
  {NoStop}%
\bibitem [{\citenamefont {Lootens}\ \emph {et~al.}(2005)\citenamefont
  {Lootens}, \citenamefont {van Damme}, \citenamefont {H{\'e}mar},\ and\
  \citenamefont {H{\'e}braud}}]{Lootens_2005}%
  \BibitemOpen
  \bibfield  {author} {\bibinfo {author} {\bibfnamefont {D.}~\bibnamefont
  {Lootens}}, \bibinfo {author} {\bibfnamefont {H.}~\bibnamefont {van Damme}},
  \bibinfo {author} {\bibfnamefont {Y.}~\bibnamefont {H{\'e}mar}}, \ and\
  \bibinfo {author} {\bibfnamefont {P.}~\bibnamefont {H{\'e}braud}},\
  }\href@noop {} {\bibfield  {journal} {\bibinfo  {journal} {Phys. Rev. Lett.}\
  }\textbf {\bibinfo {volume} {95}},\ \bibinfo {pages} {268302} (\bibinfo
  {year} {2005})}\BibitemShut {NoStop}%
\bibitem [{\citenamefont {Hsiao}\ \emph {et~al.}(2017)\citenamefont {Hsiao},
  \citenamefont {Jamali}, \citenamefont {Glynos}, \citenamefont {Green},
  \citenamefont {Larson},\ and\ \citenamefont {Solomon}}]{Hsiao_2017}%
  \BibitemOpen
  \bibfield  {author} {\bibinfo {author} {\bibfnamefont {L.~C.}\ \bibnamefont
  {Hsiao}}, \bibinfo {author} {\bibfnamefont {S.}~\bibnamefont {Jamali}},
  \bibinfo {author} {\bibfnamefont {E.}~\bibnamefont {Glynos}}, \bibinfo
  {author} {\bibfnamefont {P.~F.}\ \bibnamefont {Green}}, \bibinfo {author}
  {\bibfnamefont {R.~G.}\ \bibnamefont {Larson}}, \ and\ \bibinfo {author}
  {\bibfnamefont {M.~J.}\ \bibnamefont {Solomon}},\ }\href@noop {} {\bibfield
  {journal} {\bibinfo  {journal} {Phys. Rev. Lett.}\ }\textbf {\bibinfo
  {volume} {119}},\ \bibinfo {pages} {158001} (\bibinfo {year}
  {2017})}\BibitemShut {NoStop}%
\bibitem [{\citenamefont {Hsu}\ \emph {et~al.}(2018)\citenamefont {Hsu},
  \citenamefont {Ramakrishna}, \citenamefont {Zanini}, \citenamefont
  {Spencer},\ and\ \citenamefont {Isa}}]{Hsu_2018}%
  \BibitemOpen
  \bibfield  {author} {\bibinfo {author} {\bibfnamefont {C.-P.}\ \bibnamefont
  {Hsu}}, \bibinfo {author} {\bibfnamefont {S.~N.}\ \bibnamefont
  {Ramakrishna}}, \bibinfo {author} {\bibfnamefont {M.}~\bibnamefont {Zanini}},
  \bibinfo {author} {\bibfnamefont {N.~D.}\ \bibnamefont {Spencer}}, \ and\
  \bibinfo {author} {\bibfnamefont {L.}~\bibnamefont {Isa}},\ }\href@noop {}
  {\bibfield  {journal} {\bibinfo  {journal} {Proc. Nat. Acad. Sci.}\ }
  (\bibinfo {year} {2018})}\BibitemShut {NoStop}%
\bibitem [{\citenamefont {James}\ \emph {et~al.}(2018)\citenamefont {James},
  \citenamefont {Han}, \citenamefont {de~la Cruz}, \citenamefont {Jureller},\
  and\ \citenamefont {Jaeger}}]{james2018interparticle}%
  \BibitemOpen
  \bibfield  {author} {\bibinfo {author} {\bibfnamefont {N.~M.}\ \bibnamefont
  {James}}, \bibinfo {author} {\bibfnamefont {E.}~\bibnamefont {Han}}, \bibinfo
  {author} {\bibfnamefont {R.~A.~L.}\ \bibnamefont {de~la Cruz}}, \bibinfo
  {author} {\bibfnamefont {J.}~\bibnamefont {Jureller}}, \ and\ \bibinfo
  {author} {\bibfnamefont {H.~M.}\ \bibnamefont {Jaeger}},\ }\href@noop {}
  {\bibfield  {journal} {\bibinfo  {journal} {Nat. Mater.}\ }\textbf {\bibinfo
  {volume} {17}},\ \bibinfo {pages} {965} (\bibinfo {year} {2018})}\BibitemShut
  {NoStop}%
\bibitem [{\citenamefont {Estrada}\ \emph {et~al.}(2011)\citenamefont
  {Estrada}, \citenamefont {Az{\'e}ma}, \citenamefont {Radjai},\ and\
  \citenamefont {Taboada}}]{Estrada_2011}%
  \BibitemOpen
  \bibfield  {author} {\bibinfo {author} {\bibfnamefont {N.}~\bibnamefont
  {Estrada}}, \bibinfo {author} {\bibfnamefont {E.}~\bibnamefont {Az{\'e}ma}},
  \bibinfo {author} {\bibfnamefont {F.}~\bibnamefont {Radjai}}, \ and\ \bibinfo
  {author} {\bibfnamefont {A.}~\bibnamefont {Taboada}},\ }\href@noop {}
  {\bibfield  {journal} {\bibinfo  {journal} {Phys. Rev. E}\ }\textbf {\bibinfo
  {volume} {84}},\ \bibinfo {pages} {011306} (\bibinfo {year}
  {2011})}\BibitemShut {NoStop}%
\bibitem [{\citenamefont {Ai}\ \emph {et~al.}(2011)\citenamefont {Ai},
  \citenamefont {Chen}, \citenamefont {Rotter},\ and\ \citenamefont
  {Ooi}}]{ai2011assessment}%
  \BibitemOpen
  \bibfield  {author} {\bibinfo {author} {\bibfnamefont {J.}~\bibnamefont
  {Ai}}, \bibinfo {author} {\bibfnamefont {J.-F.}\ \bibnamefont {Chen}},
  \bibinfo {author} {\bibfnamefont {J.~M.}\ \bibnamefont {Rotter}}, \ and\
  \bibinfo {author} {\bibfnamefont {J.~Y.}\ \bibnamefont {Ooi}},\ }\href@noop
  {} {\bibfield  {journal} {\bibinfo  {journal} {Powder Technology}\ }\textbf
  {\bibinfo {volume} {206}},\ \bibinfo {pages} {269} (\bibinfo {year}
  {2011})}\BibitemShut {NoStop}%
\bibitem [{\citenamefont {Dominik}\ and\ \citenamefont
  {Tielens}(1995)}]{Dominik_1995}%
  \BibitemOpen
  \bibfield  {author} {\bibinfo {author} {\bibfnamefont {C.}~\bibnamefont
  {Dominik}}\ and\ \bibinfo {author} {\bibfnamefont {A.}~\bibnamefont
  {Tielens}},\ }\href@noop {} {\bibfield  {journal} {\bibinfo  {journal} {Phil.
  Mag. A}\ }\textbf {\bibinfo {volume} {72}},\ \bibinfo {pages} {783} (\bibinfo
  {year} {1995})}\BibitemShut {NoStop}%
\bibitem [{\citenamefont {Marshall}\ and\ \citenamefont
  {Li}(2014)}]{Marshall_2014}%
  \BibitemOpen
  \bibfield  {author} {\bibinfo {author} {\bibfnamefont {J.~S.}\ \bibnamefont
  {Marshall}}\ and\ \bibinfo {author} {\bibfnamefont {S.}~\bibnamefont {Li}},\
  }\href@noop {} {\emph {\bibinfo {title} {Adhesive particle flow}}}\ (\bibinfo
   {publisher} {Cambridge University Press},\ \bibinfo {year}
  {2014})\BibitemShut {NoStop}%
\bibitem [{\citenamefont {James}\ \emph {et~al.}(2019)\citenamefont {James},
  \citenamefont {Hsu}, \citenamefont {Spencer}, \citenamefont {Jaeger},\ and\
  \citenamefont {Isa}}]{James_2019}%
  \BibitemOpen
  \bibfield  {author} {\bibinfo {author} {\bibfnamefont {N.~M.}\ \bibnamefont
  {James}}, \bibinfo {author} {\bibfnamefont {C.-P.}\ \bibnamefont {Hsu}},
  \bibinfo {author} {\bibfnamefont {N.~D.}\ \bibnamefont {Spencer}}, \bibinfo
  {author} {\bibfnamefont {H.~M.}\ \bibnamefont {Jaeger}}, \ and\ \bibinfo
  {author} {\bibfnamefont {L.}~\bibnamefont {Isa}},\ }\href@noop {} {\bibfield
  {journal} {\bibinfo  {journal} {J. Phys. Chem. Lett.}\ }\textbf {\bibinfo
  {volume} {10}},\ \bibinfo {pages} {1663} (\bibinfo {year}
  {2019})}\BibitemShut {NoStop}%
\bibitem [{\citenamefont {Neuville}\ \emph {et~al.}(2012)\citenamefont
  {Neuville}, \citenamefont {Bossis}, \citenamefont {Persello}, \citenamefont
  {Volkova}, \citenamefont {Boustingorry},\ and\ \citenamefont
  {Mosquet}}]{Neuville_2012}%
  \BibitemOpen
  \bibfield  {author} {\bibinfo {author} {\bibfnamefont {M.}~\bibnamefont
  {Neuville}}, \bibinfo {author} {\bibfnamefont {G.}~\bibnamefont {Bossis}},
  \bibinfo {author} {\bibfnamefont {J.}~\bibnamefont {Persello}}, \bibinfo
  {author} {\bibfnamefont {O.}~\bibnamefont {Volkova}}, \bibinfo {author}
  {\bibfnamefont {P.}~\bibnamefont {Boustingorry}}, \ and\ \bibinfo {author}
  {\bibfnamefont {M.}~\bibnamefont {Mosquet}},\ }\href@noop {} {\bibfield
  {journal} {\bibinfo  {journal} {J. Rheol.}\ }\textbf {\bibinfo {volume}
  {56}},\ \bibinfo {pages} {435} (\bibinfo {year} {2012})}\BibitemShut
  {NoStop}%
\bibitem [{\citenamefont {Hsiao}\ and\ \citenamefont
  {Pradeep}(2019)}]{hsiao2019experimental}%
  \BibitemOpen
  \bibfield  {author} {\bibinfo {author} {\bibfnamefont {L.~C.}\ \bibnamefont
  {Hsiao}}\ and\ \bibinfo {author} {\bibfnamefont {S.}~\bibnamefont
  {Pradeep}},\ }\href@noop {} {\bibfield  {journal} {\bibinfo  {journal} {Curr.
  Opin. Colloid Interface Sci.}\ } (\bibinfo {year} {2019})}\BibitemShut
  {NoStop}%
\bibitem [{\citenamefont {Guy}\ \emph {et~al.}(2018)\citenamefont {Guy},
  \citenamefont {Richards}, \citenamefont {Hodgson}, \citenamefont {Blanco},\
  and\ \citenamefont {Poon}}]{Guy_2018}%
  \BibitemOpen
  \bibfield  {author} {\bibinfo {author} {\bibfnamefont {B.~M.}\ \bibnamefont
  {Guy}}, \bibinfo {author} {\bibfnamefont {J.}~\bibnamefont {Richards}},
  \bibinfo {author} {\bibfnamefont {D.}~\bibnamefont {Hodgson}}, \bibinfo
  {author} {\bibfnamefont {E.}~\bibnamefont {Blanco}}, \ and\ \bibinfo {author}
  {\bibfnamefont {W.~C.~K.}\ \bibnamefont {Poon}},\ }\href@noop {} {\bibfield
  {journal} {\bibinfo  {journal} {Phys. Rev. Lett.}\ }\textbf {\bibinfo
  {volume} {121}},\ \bibinfo {pages} {128001} (\bibinfo {year}
  {2018})}\BibitemShut {NoStop}%
\bibitem [{Note3()}]{Note3}%
  \BibitemOpen
  \bibinfo {note} {Indeed the theory with rolling friction was recently
  discussed in \protect \citep {Mari_2019})}\BibitemShut {NoStop}%
\bibitem [{\citenamefont {Luding}(2008)}]{Luding_2008}%
  \BibitemOpen
  \bibfield  {author} {\bibinfo {author} {\bibfnamefont {S.}~\bibnamefont
  {Luding}},\ }\href@noop {} {\bibfield  {journal} {\bibinfo  {journal} {Gran.
  Matt.}\ }\textbf {\bibinfo {volume} {10}},\ \bibinfo {pages} {235} (\bibinfo
  {year} {2008})}\BibitemShut {NoStop}%
\bibitem [{Not()}]{NoteX}%
  \BibitemOpen
  \href@noop {} {}\bibinfo {note} {See Supplemental Material for
  details}\BibitemShut {NoStop}%
\bibitem [{\citenamefont {Seto}\ and\ \citenamefont
  {Giusteri}(2018)}]{Seto_2018}%
  \BibitemOpen
  \bibfield  {author} {\bibinfo {author} {\bibfnamefont {R.}~\bibnamefont
  {Seto}}\ and\ \bibinfo {author} {\bibfnamefont {G.~G.}\ \bibnamefont
  {Giusteri}},\ }\href@noop {} {\bibfield  {journal} {\bibinfo  {journal} {J.
  Fluid Mech.}\ }\textbf {\bibinfo {volume} {857}},\ \bibinfo {pages} {200}
  (\bibinfo {year} {2018})}\BibitemShut {NoStop}%
\bibitem [{\citenamefont {Royer}\ \emph {et~al.}(2016)\citenamefont {Royer},
  \citenamefont {Blair},\ and\ \citenamefont {Hudson}}]{Royer_2016}%
  \BibitemOpen
  \bibfield  {author} {\bibinfo {author} {\bibfnamefont {J.~R.}\ \bibnamefont
  {Royer}}, \bibinfo {author} {\bibfnamefont {D.~L.}\ \bibnamefont {Blair}}, \
  and\ \bibinfo {author} {\bibfnamefont {S.~D.}\ \bibnamefont {Hudson}},\
  }\href@noop {} {\bibfield  {journal} {\bibinfo  {journal} {Phys. Rev. Lett.}\
  }\textbf {\bibinfo {volume} {116}},\ \bibinfo {pages} {188301} (\bibinfo
  {year} {2016})}\BibitemShut {NoStop}%
\bibitem [{\citenamefont {Tanner}\ and\ \citenamefont
  {Dai}(2016)}]{Tanner_2016}%
  \BibitemOpen
  \bibfield  {author} {\bibinfo {author} {\bibfnamefont {R.~I.}\ \bibnamefont
  {Tanner}}\ and\ \bibinfo {author} {\bibfnamefont {S.}~\bibnamefont {Dai}},\
  }\href@noop {} {\bibfield  {journal} {\bibinfo  {journal} {J. Rheol.}\
  }\textbf {\bibinfo {volume} {60}},\ \bibinfo {pages} {809} (\bibinfo {year}
  {2016})}\BibitemShut {NoStop}%
\bibitem [{Note4()}]{Note4}%
  \BibitemOpen
  \bibinfo {note} {If $N_1$ data are also available, we may be able to
  determine $\mu _s$ and $\mu _r$ uniquely.}\BibitemShut {Stop}%
\bibitem [{\citenamefont {Gameiro}\ \emph {et~al.}(2020)\citenamefont
  {Gameiro}, \citenamefont {Singh}, \citenamefont {Kondic}, \citenamefont
  {Mischaikow},\ and\ \citenamefont {Morris}}]{gameiro2019interaction}%
  \BibitemOpen
  \bibfield  {author} {\bibinfo {author} {\bibfnamefont {M.}~\bibnamefont
  {Gameiro}}, \bibinfo {author} {\bibfnamefont {A.}~\bibnamefont {Singh}},
  \bibinfo {author} {\bibfnamefont {L.}~\bibnamefont {Kondic}}, \bibinfo
  {author} {\bibfnamefont {K.}~\bibnamefont {Mischaikow}}, \ and\ \bibinfo
  {author} {\bibfnamefont {J.~F.}\ \bibnamefont {Morris}},\ }\href@noop {}
  {\bibfield  {journal} {\bibinfo  {journal} {Phys. Rev. Fluids}\ }\textbf
  {\bibinfo {volume} {5}},\ \bibinfo {pages} {034307} (\bibinfo {year}
  {2020})}\BibitemShut {NoStop}%
\bibitem [{\citenamefont {Lois}\ \emph {et~al.}(2007)\citenamefont {Lois},
  \citenamefont {Lema{\^\i}tre},\ and\ \citenamefont
  {Carlson}}]{lois2007spatial}%
  \BibitemOpen
  \bibfield  {author} {\bibinfo {author} {\bibfnamefont {G.}~\bibnamefont
  {Lois}}, \bibinfo {author} {\bibfnamefont {A.}~\bibnamefont {Lema{\^\i}tre}},
  \ and\ \bibinfo {author} {\bibfnamefont {J.~M.}\ \bibnamefont {Carlson}},\
  }\href@noop {} {\bibfield  {journal} {\bibinfo  {journal} {Phys. Rev. E}\
  }\textbf {\bibinfo {volume} {76}},\ \bibinfo {pages} {021302} (\bibinfo
  {year} {2007})}\BibitemShut {NoStop}%
\bibitem [{\citenamefont {Radjai}\ \emph {et~al.}(1998)\citenamefont {Radjai},
  \citenamefont {Wolf}, \citenamefont {Jean},\ and\ \citenamefont
  {Moreau}}]{Radjai_1998}%
  \BibitemOpen
  \bibfield  {author} {\bibinfo {author} {\bibfnamefont {F.}~\bibnamefont
  {Radjai}}, \bibinfo {author} {\bibfnamefont {D.~E.}\ \bibnamefont {Wolf}},
  \bibinfo {author} {\bibfnamefont {M.}~\bibnamefont {Jean}}, \ and\ \bibinfo
  {author} {\bibfnamefont {J.-J.}\ \bibnamefont {Moreau}},\ }\href@noop {}
  {\bibfield  {journal} {\bibinfo  {journal} {Phys. Rev. Lett.}\ }\textbf
  {\bibinfo {volume} {80}},\ \bibinfo {pages} {61} (\bibinfo {year}
  {1998})}\BibitemShut {NoStop}%
\bibitem [{\citenamefont {Cwalina}\ \emph {et~al.}(2016)\citenamefont
  {Cwalina}, \citenamefont {Harrison},\ and\ \citenamefont
  {Wagner}}]{cwalina2016rheology}%
  \BibitemOpen
  \bibfield  {author} {\bibinfo {author} {\bibfnamefont {C.~D.}\ \bibnamefont
  {Cwalina}}, \bibinfo {author} {\bibfnamefont {K.~J.}\ \bibnamefont
  {Harrison}}, \ and\ \bibinfo {author} {\bibfnamefont {N.~J.}\ \bibnamefont
  {Wagner}},\ }\href@noop {} {\bibfield  {journal} {\bibinfo  {journal} {Soft
  Matter}\ }\textbf {\bibinfo {volume} {12}},\ \bibinfo {pages} {4654}
  (\bibinfo {year} {2016})}\BibitemShut {NoStop}%
\bibitem [{\citenamefont {Cwalina}\ \emph {et~al.}(2017)\citenamefont
  {Cwalina}, \citenamefont {Harrison},\ and\ \citenamefont
  {Wagner}}]{cwalina2017rheology}%
  \BibitemOpen
  \bibfield  {author} {\bibinfo {author} {\bibfnamefont {C.~D.}\ \bibnamefont
  {Cwalina}}, \bibinfo {author} {\bibfnamefont {K.~J.}\ \bibnamefont
  {Harrison}}, \ and\ \bibinfo {author} {\bibfnamefont {N.~J.}\ \bibnamefont
  {Wagner}},\ }\href@noop {} {\bibfield  {journal} {\bibinfo  {journal} {AIChE
  Journal}\ }\textbf {\bibinfo {volume} {63}},\ \bibinfo {pages} {1091}
  (\bibinfo {year} {2017})}\BibitemShut {NoStop}%
\bibitem [{\citenamefont {He}\ and\ \citenamefont {Thorpe}(1985)}]{he1985}%
  \BibitemOpen
  \bibfield  {author} {\bibinfo {author} {\bibfnamefont {H.}~\bibnamefont
  {He}}\ and\ \bibinfo {author} {\bibfnamefont {M.~F.}\ \bibnamefont
  {Thorpe}},\ }\href@noop {} {\bibfield  {journal} {\bibinfo  {journal} {Phys.
  Rev. Lett.}\ }\textbf {\bibinfo {volume} {54}},\ \bibinfo {pages} {2107}
  (\bibinfo {year} {1985})}\BibitemShut {NoStop}%
\bibitem [{\citenamefont {Singh}\ \emph {et~al.}(2015)\citenamefont {Singh},
  \citenamefont {Magnanimo}, \citenamefont {Saitoh},\ and\ \citenamefont
  {Luding}}]{Singh_2015}%
  \BibitemOpen
  \bibfield  {author} {\bibinfo {author} {\bibfnamefont {A.}~\bibnamefont
  {Singh}}, \bibinfo {author} {\bibfnamefont {V.}~\bibnamefont {Magnanimo}},
  \bibinfo {author} {\bibfnamefont {K.}~\bibnamefont {Saitoh}}, \ and\ \bibinfo
  {author} {\bibfnamefont {S.}~\bibnamefont {Luding}},\ }\href@noop {}
  {\bibfield  {journal} {\bibinfo  {journal} {New J. Phys}\ }\textbf {\bibinfo
  {volume} {17}},\ \bibinfo {pages} {043028} (\bibinfo {year}
  {2015})}\BibitemShut {NoStop}%
\bibitem [{\citenamefont {Mari}\ \emph {et~al.}(2015)\citenamefont {Mari},
  \citenamefont {Seto}, \citenamefont {Morris},\ and\ \citenamefont
  {Denn}}]{Mari_2015}%
  \BibitemOpen
  \bibfield  {author} {\bibinfo {author} {\bibfnamefont {R.}~\bibnamefont
  {Mari}}, \bibinfo {author} {\bibfnamefont {R.}~\bibnamefont {Seto}}, \bibinfo
  {author} {\bibfnamefont {J.~F.}\ \bibnamefont {Morris}}, \ and\ \bibinfo
  {author} {\bibfnamefont {M.~M.}\ \bibnamefont {Denn}},\ }\href@noop {}
  {\bibfield  {journal} {\bibinfo  {journal} {Phys. Rev. E}\ }\textbf {\bibinfo
  {volume} {91}},\ \bibinfo {pages} {052302} (\bibinfo {year}
  {2015})}\BibitemShut {NoStop}%
\end{thebibliography}%
\bibliographystyle{apsrev4-1}

%

\clearpage

\begin{widetext}

\begin{appendix}

\section*{Supplemental Material for ``Shear thickening and jamming of dense suspensions:
the \emph{roll} of friction"}
In this document we provide details about the rolling friction forces used in the simulations. 

\beginsupplement

\section*{Rolling Friction}\label{rol_fric}

In the simulation scheme used in this article, the particles interact through near-field hydrodynamic
interactions (lubrication), a conservative repulsive force, and frictional contact forces.


In this work, we follow \citet{Luding_2008} to model the contact forces including sliding and rolling frictions.
We assume two  particles having radii $a_i$ and $a_j$ having $\boldsymbol{U}^{(i)}$ and $\boldsymbol{U}^{(j)}$ as translational and
$\boldsymbol{\Omega}^{(i)}$ and $\boldsymbol{\Omega}^{(j)}$ as rotational velocities, respectively.
The contact force between two particles is active only when 
overlap $\delta^{(i,j)} \equiv a_i+a_j - |\boldsymbol{r}_i - \boldsymbol{r}_j| $ is positive.

%

The normal (volume-excluding) force $\boldsymbol{F}_{\mathrm{C,nor}}$, 
sliding-friction force $\boldsymbol{F}_{\mathrm{C,slid}}$,
sliding-friction torque $\boldsymbol{T}_{\mathrm{C,slid}}$, 
and rolling-friction torque  $\boldsymbol{T}_{\mathrm{C,roll}}$
between the two particles are obtained as:
\begin{subequations}
\begin{equation}
\boldsymbol{F}_{\mathrm{C,nor}}^{(i,j)} = k_{n}\delta^{(i,j)}\boldsymbol{n}_{ij}~,
\end{equation}
\begin{equation}
\boldsymbol{F}_{\mathrm{C,slid}}^{(i,j)} = k_{t} \boldsymbol{\xi}^{(i,j)}~,
\end{equation}
\begin{equation}
\boldsymbol{T}_{\mathrm{C,slid}}^{(i,j)} = a_i \boldsymbol{n}_{ij} 
\times \boldsymbol{F}_{\mathrm{C,slid}}^{(i,j)}~,
\end{equation}
\begin{equation}
\boldsymbol{T}_{\mathrm{C,roll}}^{(i,j)} = 
a_{ij} \boldsymbol{n}_{ij} \times \boldsymbol{F}_{\mathrm{C,roll}}^{(i,j)}~.
\end{equation}
\end{subequations}
%
%
Here, $\boldsymbol{n_{ij}} \equiv (\boldsymbol{r}_i - \boldsymbol{r}_j)/|\boldsymbol{r}_i - \boldsymbol{r}_j|$ is the unit vector that points from particle $j$ to $i$,
and $a_{ij} \equiv 2 a_i a_j/(a_i + a_j)$ is the reduced radius.
Note that $\boldsymbol{F}_{\mathrm{C,roll}}^{(i,j)}$
\begin{equation}
\boldsymbol{F}_{\mathrm{C,roll}}^{(i,j)} = k_{r} \boldsymbol{\psi}^{(i,j)}~,
\end{equation}
is a quasi-force, which is used only to compute torque, $\boldsymbol{T}_{\mathrm{C,roll}}^{(i,j)} $.
The parameters $k_n$, $k_t$ and $k_r$ are 
the normal, sliding and rolling spring constants, respectively.

The spring stretches in sliding $\boldsymbol{\xi}^{(i,j)}(t)$ 
and rolling $\boldsymbol{\psi}^{(i,j)}(t)$ modes
are given 
by the following integrals of relative velocities from 
the time $t_c$ when the contact appears:
\begin{subequations}
\begin{equation}
\boldsymbol{\xi}^{(i,j)}(t) = \int_{t_c}^{t} \boldsymbol{U}_{\mathrm{t}}^{(i,j)} dt~,
\end{equation}
\begin{equation}
\boldsymbol{\psi}^{(i,j)}(t) = \int_{t_c}^{t} \boldsymbol{U}_{\mathrm{r}}^{(i,j)} dt~,
\end{equation}
\end{subequations}
as long as the sliding and rolling frictions
to fulfill Coulomb's friction laws:
$|\boldsymbol{F}_{\mathrm{C,slid}}^{(i,j)}| \le 
\mu_s |\boldsymbol{F}_{\mathrm{C,nor}}^{(i,j)}| $ 
and $|\boldsymbol{F}_{\mathrm{C,roll}}^{(i,j)}| 
\le \mu_r |\boldsymbol{F}_{\mathrm{C,nor}}^{(i,j)}|$, 
with sliding $\mu_s$ and rolling $\mu_r$ friction coefficients.
Otherwise, the spring stretches are adjusted to 
keep the maximum values of the inequalities
(For simplicity, we do not set the dynamic friction coefficients).
The normal $\boldsymbol{U}_{\mathrm{n}}^{(i,j)}$, 
tangential $\boldsymbol{U}_{\mathrm{t}}^{(i,j)}$, 
and rolling $\boldsymbol{U}_{\mathrm{r}}^{(i,j)}$ relative velocities 
between two particles $i$ and $j$ are given by:
\begin{subequations}
\begin{equation}
\boldsymbol{U}_{\mathrm{n}}^{(i,j)} \equiv \boldsymbol{\mathsf{P}}_{n_{ij}} (\boldsymbol{U}^{(j)} - \boldsymbol{U}^{(i)})~,
\end{equation}
\begin{equation}
\boldsymbol{U}_{\mathrm{t}}^{(i,j)} \equiv \boldsymbol{\mathsf{P}}_{n_{ij}}^\prime 
\bigl[ (\boldsymbol{U}^{(j)} - \boldsymbol{U}^{(i)}) -(a_i\boldsymbol{\Omega}^{(i)} + a_j\boldsymbol{\Omega}^{(j)}) \times \boldsymbol{n}_{ij}  \bigr]~,
\end{equation}
\begin{equation}
\boldsymbol{U}_{\mathrm{r}}^{(i,j)} \equiv {a}_{{ij}} (\boldsymbol{\Omega}^{(i)} - \boldsymbol{\Omega}^{(j)}) \times \boldsymbol{n}_{ij}~,
\end{equation}
\end{subequations}
where $\boldsymbol{\mathsf{P}}_{n_{ij}} \equiv \boldsymbol{n}_{ij} \boldsymbol{n}_{ij}$ 
is the normal projection operator, 
while $\boldsymbol{\mathsf{P}}_{n_{ij}}^\prime \equiv 
\boldsymbol{\mathsf{I}} - \boldsymbol{n}_{ij} \boldsymbol{n}_{ij}$ 
is the tangential projection operator, which was introduced previously\,\cite{Mari_2014}.

%
%
%
%

%
Finally, the total contact force and torque are given by:
\begin{subequations}
\begin{equation}
\boldsymbol{F}_{\mathrm{C}}^{(i,j)} = 
\boldsymbol{F}_{\mathrm{C,nor}}^{(i,j)}  + \boldsymbol{F}_{\mathrm{C,slid}}^{(i,j)}~,
\end{equation}
\begin{equation}
\boldsymbol{T}_{\mathrm{C}}^{(i,j)} = 
a_i \boldsymbol{n}_{ij} \times \boldsymbol{F}_{\mathrm{C,slid}}
  + a_{ij}\boldsymbol{n}_{ij} \times \boldsymbol{F}_{\mathrm{C,roll}}~.
\end{equation}
\end{subequations}
%
We use spring stiffnesses such that the maximum particle overlaps do not exceed 3\% of the particle radius in order to stay
close to the rigid limit\,\citep{Singh_2015,Ness_2016}.
Note that we do not use any dashpot explicitly, but to stabilize the simulation we make use of lubrication resistance that acts as a dashpot\,\cite{Mari_2015}.

\end{appendix}
\clearpage

\end{widetext}

%
\end{document}